\documentclass[conference]{IEEEtran}

\usepackage{amsmath,amssymb}
\usepackage{graphicx}
\usepackage{cite}
\usepackage{url}
\usepackage[letterpaper,top=0.75in,bottom=1in,left=0.625in,right=0.625in]{geometry}
\usepackage{array}
\usepackage{newtxtext,newtxmath}
\usepackage{refcount}

\begin{document}
\raggedbottom

\title{Profiling and Scheduling Complex O-RAN Applications Across the 5G Edge and Cloud}
\author{\IEEEauthorblockN{Yoonjae Hwang and Bhaskar Krishnamachari}
\IEEEauthorblockA{Viterbi School of Engineering, University of Southern California\\ \texttt{\char123 yoonjae, bkrishna\char125@usc.edu}}
}
\maketitle

\begin{abstract}
The O-RAN paradigm decomposes intelligent RAN control into pipelines of interdependent AI/ML functions, including traffic prediction, signal quality estimation, and slice scheduling, that must execute across a dispersed continuum of far-edge, near-edge, and cloud resources under heterogeneous latency and bandwidth constraints. Despite the natural expression of these pipelines as Directed Acyclic Graphs (DAGs), no integrated methodology exists to profile their execution costs, map them onto dispersed infrastructure via scheduling heuristics, and validate the resulting placement under 5G cellular conditions. We present O-DAG, an end-to-end framework that closes this gap through four tightly coupled stages: (1) DagProfiler, a new open-source tool that instruments O-RAN Slice Scheduler and extracts per-task instruction counts and per-edge communication volumes; (2) a parameterized three-tier network topology encoding far-edge (DU, RIC), near-edge (edge), and cloud nodes with realistic link bandwidths; (3) an extension of the SAGA scheduling framework and (4) a custom DAG simulation module built on the MintEDGE simulator. We evaluate five scheduling algorithms (HEFT, MCT, MinMin, MaxMin, Duplex) for a slice scheduling application across various configurations spanning 5K--50K UEs, 2--20 cells, and 2--10 network slices. HEFT achieves the lowest makespan in all configurations, but scheduler rankings are workload-dependent. The SAGA--simulation gap serves as a regime diagnostic: negative gaps (up to $-1.72$\%) identify compute-dominated configurations where HEFT overestimates conservatively, while a positive gap (+0.64\%) at low slice counts exposes a communication-bound regime where bandwidth contention exceeds the scheduling model's assumptions. All artifacts are released for reproducibility.
\end{abstract}

% NOTE: IEEEtran uses IEEEkeywords; keeping the user's "Key Terms" text as regular content is not recommended.

\section{Introduction}

The evolution toward 5G and 6G architectures demands a shift from static network configurations to dynamic, AI-driven orchestration [1]--[3]. Network slicing---the dynamic partitioning of radio resources across heterogeneous service requirements---is central to this vision. The O-RAN Alliance’s architecture enables slice scheduling via xApps and rApps hosted on the RAN Intelligent Controller (RIC) [1], while the recently proposed dApps concept extends this control to sub-10 ms timescales at the CU/DU level [4]. Realizing an effective slicing scheduler, however, is no longer a localized task: it has become a distributed workflow of interdependent functions---traffic prediction, signal quality estimation, inter- and intra-slice scheduling---that must execute across a heterogeneous compute continuum spanning the far edge, near edge, and cloud [5].

Consider the RAN slicing scheduler described by Ananthanarayanan et al.\ [5]: buffer status reports feed a traffic predictor, sounding reference signals feed a signal quality predictor, and both inform an inter-slice scheduler that operates at coarse timescales and a real-time intra-slice scheduler at the far edge. These functional blocks form a natural Directed Acyclic Graph (DAG) with distinct compute profiles and data dependencies. Profiling the execution cost of each task and the data volume along each edge is a prerequisite for placement, yet the community lacks integrated methodologies to perform this profiling, translate results into scheduling decisions, and validate them under 5G network conditions.

This gap persists because existing tools address only fragments of the problem. The canonical HEFT algorithm [6] schedules DAGs on heterogeneous processors but assumes deterministic costs and ignores communication contention. Contention-aware [7] and uncertainty-aware [8], [9] schedulers model shared-link and execution-time variability, respectively, but target traditional cloud settings. DIME [10] demonstrates that accurate bundle-level profiling is critical for distributed inference, but does not integrate with DAG scheduling. The SAGA benchmarking framework [11] provides 17 scheduling algorithms and adversarial analysis, but lacks O-RAN-specific task graphs. Meanwhile, simulation platforms like MintEDGE [12] and Colosseum [13] offer realistic 5G evaluation environments, but have not been coupled with DAG profiling or dispersed scheduling.

This operational complexity is compounded by the trade-offs of dispersed computing---the coordinated execution across far edge ($<1$ ms, $>100$ Gbps), near edge (1--10 ms, $<100$ Gbps), and cloud ($>10$ ms, $<10$ Gbps) tiers [5]. The cloud offers GPU abundance but prohibitive transport latency; the far edge minimizes latency but faces severe resource contention. As deployments scale from single-cell to metropolitan topologies, the dominant latency contributor shifts from compute to transport---a phenomenon our framework is designed to expose.

To bridge these gaps, we present O-DAG (Open-RAN Directed Acyclic Graph framework), an end-to-end framework for profiling, scheduling, and evaluating O-RAN slicing workflows in four stages: (1) we formalize the slice scheduler DAG using DagProfiler, extracting per-task compute costs and per-edge communication volumes; (2) we define a heterogeneous network topology capturing the three-tier continuum; (3) we apply SAGA’s HEFT-based scheduler [6], [11] to derive task-to-node mappings; and (4) we instantiate these mappings in MintEDGE [12] to quantify performance under 5G cellular topologies. Our evaluation decomposes end-to-end delay into compute, transport, and contention components to reveal critical-path shifts as the network scales.

The main contributions are as follows:
\begin{itemize}
  \item \textbf{Reusable O-RAN application DAG benchmark.} We formalize a concrete 5G/O-RAN operations workflow as a DAG (task semantics + dependencies), and provide a profiling schema capturing per-task compute costs and per-edge communication volumes.
  \item \textbf{End-to-end toolchain from DAG scheduling to realistic 5G execution (SAGA $\rightarrow$ MintEdge).} We take a profiled DAG, compute dispersed placements via SAGA scheduling, and instantiate these placements in MintEdge to measure realized compute + communication time on cellular topologies.
  \item \textbf{Slice- and topology-aware scaling characterization with clean latency decomposition.} We characterize how latency scales with network size, number of slices, and UE load, and decompose end-to-end delay into compute, transport, and contention components to explain critical-path shifts across tiers.
  \item \textbf{Reproducible artifacts and a generalizable methodology.} We package the DAG spec, profiling format, and experiment configurations, and outline how the same pipeline applies to other O-RAN applications with minimal changes.
\end{itemize}

The remainder of this paper is organized as follows. Section II surveys related work across four domains: O-RAN architecture and intelligence, DAG modeling and profiling, DAG-based scheduling in edge computing, and dispersed edge–cloud scheduling and 5G simulation. Section III formalizes the RAN slicing scheduler as an eight-task DAG and describes the DagProfiler instrumentation methodology. Section IV details the end-to-end O-DAG pipeline. Section V describes the experimental setup, parameter space, profiled DAG costs, and evaluation metrics. Section VI presents and analyzes results across three experimental axes simulation gap. Section VII concludes the paper and outlines directions for future work.

\section{Related Work}

\subsection{O-RAN Architecture and Intelligence}
The O-RAN Alliance~[1] established the reference architecture for intelligent, disaggregated RANs, with ETSI TS 103 982~[2] formalizing the management layer. Applications operate at three timescales: non-RT ($\geq 1$ s), near-RT (10 ms--1 s), and RT ($< 10$ ms)~[1]. Schmidt et al.~[14] introduced FlexRIC, an ultra-lean SDK for composing SD-RAN controllers with modular service models, achieving round-trip latencies an order of magnitude lower than the reference O-RAN RIC. D'Oro et al.~[4] proposed dApps at the CU/DU level, reducing E2 interface overhead by up to 3.57$\times$ while enabling sub-10 ms beam management and scheduling control on the Colosseum testbed~[13]. Ngo et al.~[3] survey the RIC split and open-source efforts. Ananthanarayanan et al.~[5] formalized the distributed AI platform for 6G, presenting the slicing scheduler and anomaly detection pipelines as graphs of AI blocks placed across three tiers by an orchestrator. Their contribution is architectural; ours operationalizes it with profiling, scheduling, and simulation.

\subsection{DAG Modeling, Benchmarking, and Profiling}
DAG scheduling on heterogeneous platforms is NP-hard and not constant-factor approximable~[11]. Kliazovich et al.~[15] proposed CA-DAG, separating compute and communication nodes---conceptually aligned with our profiling schema but targeting cloud datacenters. Coleman and Krishnamachari~[11] built the SAGA framework with 17 algorithms, 16 datasets, and the PISA adversarial analysis method, revealing that algorithms appearing similar on benchmarks diverge on adversarial instances. Their parameterized scheduler~[16] further isolates which algorithmic components drive performance. SAGA serves as our scheduling engine, but its datasets lack O-RAN task graphs---a gap DagProfiler fills. Viramontes and Davoodi~[10] showed with DIME that naive per-layer profiling causes sub-optimal integer linear program (ILP) solutions in distributed inference, motivating our emphasis on empirical bundle-level cost capture.

\subsection{Dag-Based Task Scheduling in Edge Computing}
Recent literature has explored DAG scheduling across multi-tier edge environments. Various frameworks have been proposed to optimize execution delay, communication costs, and response times across diverse architectures for domains ranging from video analytics~[17] and 3D vision~[18] to IoT~[19] and general machine learning~[20]. However, while these frameworks effectively schedule complex DAGs in generalized edge environments, they do not integrate with O-RAN-specific profiling or cellular network simulation. Our framework closes this gap by coupling empirical bundle-level cost capture via DagProfiler with 5G cellular topologies. A detailed feature comparison of these existing edge schedulers against O-DAG is illustrated in Table~\ref{tab:edge-dag-frameworks}.

\begin{table*}[t]
  \centering
  \caption{Comparison of DAG Scheduling Frameworks in Edge and O-RAN Environments}
  \label{tab:edge-dag-frameworks}
  \small
  \begingroup
  \setlength{\tabcolsep}{3pt}
  \begin{tabular}{@{}>{\raggedright\arraybackslash}p{0.12\textwidth} >{\raggedright\arraybackslash}p{0.12\textwidth} >{\raggedright\arraybackslash}p{0.16\textwidth} >{\raggedright\arraybackslash}p{0.24\textwidth} >{\raggedright\arraybackslash}p{0.22\textwidth}@{}}
    \hline
    Framework & Target Domain & Compute Topology & Scheduling Approach & Optimization Objective \\
    \hline
    LAVEA~[17] & Video Analytics & Client, Edge, Cloud & Two-stage job-shop model with topological ordering & Minimize makespan/response time \\
    Hetero-Edge~[18] & Real-time Vision & Heterogeneous edge (CPUs, GPUs) & Latency-aware task scheduling + proportional stream grouping & End-to-end latency reduction and straggler avoidance \\
    DCC~[19] & IoT & Device, Cloudlet, Cloud & Greedy Task Graph Partition & Minimize communication cost and execution delay \\
    M-TEC~[20] & General ML & End devices (P2P), edge servers, cloud & Decentralized greedy scheduling & Joint optimization of latency, failure probability, and cost \\
    O-DAG Framework & O-RAN slicing workflows & Far edge, near edge, cloud & DagProfiler with extended SAGA framework & Makespan minimization across 5G topologies \\
    \hline
  \end{tabular}
  \endgroup
\end{table*}

\subsection{Dispersed Scheduling for Edge--Cloud}
HEFT~[6] remains the dominant heuristic for heterogeneous DAG scheduling, ranking tasks by upward rank and greedily assigning them to the earliest-finishing processor. Wu et al.~[7] extended this paradigm with ELSH, a contention-aware heuristic that co-schedules data transfers alongside tasks, yielding substantial makespan improvements for high-compute-to-communication workflows such as CyberShake and Montage. Chen et al.~[8] addressed execution-time uncertainty via a scheduling architecture that prevents uncertainty propagation across tasks, while Genez et al.~[9] handled dynamic bandwidth fluctuations for workflow ensembles---both relevant to O-RAN's variable RF and transport conditions. De Queiroz et al.~[21] proposed FlexDO for MEC DAG offloading, demonstrating that the one-climb policy does not hold for general DAGs and achieving near-optimal makespans on realistic Alibaba traces. None of these works integrates with O-RAN-specific profiling or cellular network simulation.

\subsection{5G Simulation and Latency Evaluation}
MintEDGE~[12] provides a multi-tier 5G edge simulation framework with configurable base stations, backhaul links, co-located servers, and realistic user mobility via OpenStreetMaps. Its delay model decomposes latency into uplink, routing, compute, return-routing, and downlink components, aligning naturally with our three-tier topology. We adopt MintEDGE over hardware testbeds like Colosseum~[13] to enable large-scale parametric studies. Colosseum's 256-SDR emulation platform remains the gold standard for protocol-level validation of xApp/dApp behavior, and our MintEDGE results could be cross-validated there in future work. For latency observability, Sigelman et al.~[22] established span-based distributed tracing with Dapper, decomposing end-to-end latency into hierarchical process-level spans. We adapt this concept to the O-RAN context: each DAG task corresponds to a span, and the critical path through the span tree determines slice scheduler latency, decomposed into compute, transport, and contention components.

The works surveyed above each address an isolated fragment of the problem O-DAG targets. O-DAG closes this gap by being the first framework to unify all four concerns—empirical O-RAN application profiling, heterogeneous DAG scheduling, node-pinning for physical-layer locality, and 5G cellular simulation—into a single end-to-end pipeline. The result is a methodology that not only produces scheduling decisions but validates them under realistic network conditions and exposes the compute-versus-communication regime shifts that govern O-RAN workflow performance.

\section{O-RAN Application as a Task Graph}
\subsection{Slicing Scheduler DAG}
We derive our task graph from the RAN slicing scheduler of Ananthanarayanan et al.~[5], instantiated as a concrete DAG of eight tasks (T0--T7), illustrated in Fig.~1 and summarized in Table~II. T0 (Collect State) is the pinned entry task at the far edge, fanning out to two parallel ML-inference branches: T1 (Traffic Predict) $\rightarrow$ T3 (Inter Slice Budget) for coarse-grained inter-slice PRB allocation, and T2 (Link Quality Map) $\rightarrow$ T4 (Build UE Weights) for per-UE intra-slice weights. Both branches converge at T5 (Intra Slice Schedule), which must complete within the sub-millisecond MAC scheduling window at the far edge, followed by terminal tasks T6 and T7 for enforcement and feedback. This fork--join structure creates two critical paths---T0$\rightarrow$T1$\rightarrow$T3$\rightarrow$T5$\rightarrow$T6$\rightarrow$T7 and T0$\rightarrow$T2$\rightarrow$T4$\rightarrow$T5$\rightarrow$T6$\rightarrow$T7---whose dominance depends on per-task compute profiles and tier placement: the ML-heavy T1 and T2 favor GPU-equipped cloud nodes, while T5's strict latency constraint pulls the intra-slice path toward the far edge. We implemented this scheduler as an open-source, profiling-instrumented Python application with configurable UE, cell, and slice parameters.\footnote{Slicing Scheduler: \url{https://github.com/ANRGUSC/oran-slice-scheduler}\label{fn:oransched}}

\begin{figure*}[t]
  \centering
  \fbox{\includegraphics[width=\dimexpr\textwidth-2\fboxsep-2\fboxrule\relax]{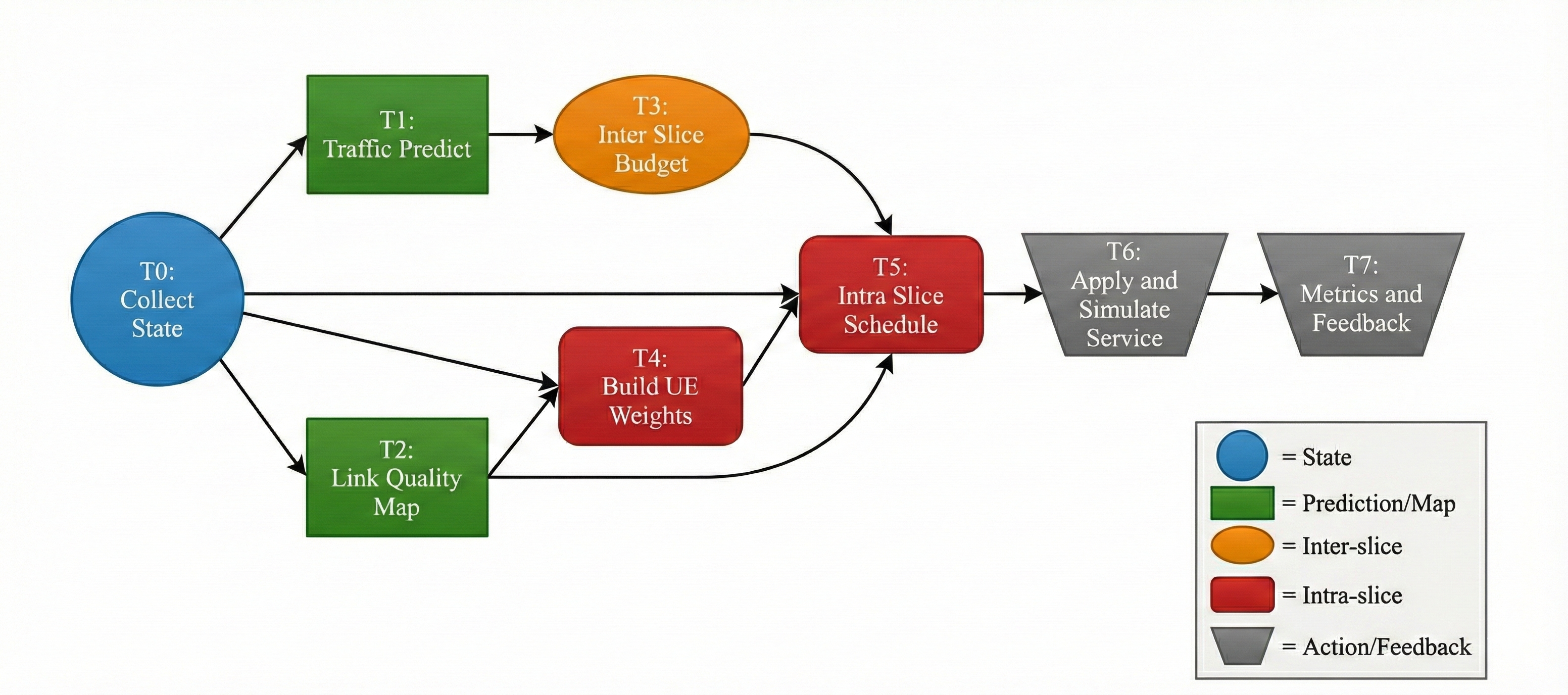}}
  \caption{Slicing scheduler DAG.}
  \label{fig:slicing-dag}
\end{figure*}

\begin{table}[t]
  \centering
  \caption{Slicing scheduler DAG task summary.}
  \label{tab:slicing-dag-tasks}
  % Placeholder: add Table I here
  \begin{tabular}{l l l l}
    \hline
    Task & Description & Role & Depends On \\
    \hline
    T0 & Collect State & State & --- (entry) \\
    T1 & Traffic Predict & Predict & T0 \\
    T2 & Link Quality Map & Predict & T0 \\
    T3 & Inter Slice Budget & Inter-slice & T1 \\
    T4 & Build UE Weights & Intra-slice & T2 \\
    T5 & Intra Slice Schedule & Intra-slice & T0, T2, T3, T4 \\
    T6 & Apply and Simulate Service & Action & T5 \\
    T7 & Metrics and Feedback & Feedback & T6 \\
    \hline
  \end{tabular}
\end{table}

\subsection{Profiling with DagProfiler}
To translate the abstract DAG into a schedulable task graph, each node and edge must be annotated with empirical cost measurements. We developed DagProfiler,\footnote{Open-source on PyPI and at \url{https://github.com/ANRGUSC/dagprofiler}} a Python framework for authoring profiler-compatible DAG tasks with automatic compute and communication profiling. DagProfiler addresses the profiling-accuracy concern raised by DIME~[10]: rather than relying on analytical cost estimates that compound errors across task boundaries, it captures end-to-end empirical measurements including framework overhead, memory allocation, and model inference.

DagProfiler operates in three phases. In the instrumentation phase, the developer wraps each task function with a lightweight profiling decorator that records wall-clock execution time, peak memory, and the serialized size (in bytes) of all inputs and outputs. In the execution phase, the instrumented DAG runs on a reference platform across multiple invocations to obtain stable estimates. In the export phase, DagProfiler produces two annotation sets:
(1) Per-task compute cost $c(t_i)$: the median execution time (seconds) of task $t_i$, representing the node weight.
(2) Per-edge communication volume $d(e_{i,j})$: the serialized data size (bytes) transferred from $t_i$ to $t_j$. Combined with link bandwidth from the network topology, this yields HEFT's communication time~[6].

The exported profile is formatted as a SAGA-compatible task graph specification~[11], enabling direct ingestion by any of SAGA's 17 scheduling algorithms without conversion. While we demonstrate DagProfiler on the slicing scheduler, its schema generalizes to any DAG-structured O-RAN application (e.g., the anomaly detection pipeline in~[5]).

\section{End-to-End Implementation Pipeline}
The O-DAG framework is organized as a four-stage sequential pipeline in which each stage consumes the outputs of the previous and produces structured artifacts that serve as inputs to the next. In the first stage, DagProfiler instruments the target O-RAN application and produces a JSON-serialized task graph annotated with per-task instruction counts and per-edge communication volumes in bits. In the second stage, a parameterized network topology definition encodes the three-tier compute continuum—far-edge DU nodes, near-edge EDGE and RIC nodes, and a GPU-equipped Cloud node—as a weighted graph with per-link bandwidth values, compatible with both the scheduling and simulation stages. In the third stage, the extended SAGA scheduling engine ingests the profiled task graph and topology graph, applies a chosen heuristic (e.g., HEFT) subject to node-pinning constraints, and outputs a task-to-node mapping with predicted start and finish times as a second JSON artifact. In the fourth stage, the custom MintEDGE DAG simulation module replays this mapping under 5G cellular conditions, enforcing dependency ordering, modeling bandwidth sharing across concurrent transfers, and decomposing each task's realized latency into uplink, compute, and downlink components. The data flow is strictly unidirectional: DagProfiler → topology definition → extended SAGA → MintEDGE, with no feedback between stages. This linear structure makes the pipeline fully reproducible and allows any individual stage to be replaced or extended independently.

\subsection{Stage 1: Task Graph Profiling}
As described in Section~III-B, DagProfiler instruments the slicing scheduler application and exports per-task compute costs (in number of instructions) and per-edge communication volumes (in bits). For the eight-task DAG with 10{,}000 UEs, 7 cells, and 5 slices, profiling yields instruction counts ranging from 54{,}150 (T1, Traffic Predict) to 7{,}720{,}000 (T2, Link Quality Map). The largest data transfers are T0$\rightarrow$T1 at 4.4~Mbit (state vector to traffic predictor) and T0$\rightarrow$T4 at 4.24~Mbit (state to UE weight builder). The profiled DAG is serialized as a JSON file specifying nodes with instruction counts and directed edges with bit volumes, serving as the first input to the scheduling stage.

\subsection{Stage 2: Network Topology Definition}
The network topology, derived from Ananthanarayanan et al.\ [5], encodes the three-tier dispersed infrastructure as a weighted graph compatible with both SAGA and MintEDGE. Nodes represent compute sites, each annotated with a speed factor $s(v)$ that multiplies the base processing rate (instructions per second). Edges represent communication links annotated with bandwidth in Gbps. We define four node types:
\begin{itemize}
  \item \textbf{DU nodes} ($s=1$): far-edge Distributed Units co-located with base stations. The number of DUs is configurable, corresponding to the cell count in the deployment.
  \item \textbf{EDGE node} ($s=3$): a near-edge aggregation server connected to all DUs via 100~Gbps fronthaul links.
  \item \textbf{RIC node} ($s=3$): the near-RT RAN Intelligent Controller, connected to each DU at 1~Gbps (E2 interface) and to EDGE at 50~Gbps.
  \item \textbf{Cloud node} ($s=6$): a GPU-equipped data center connected to RIC at 10~Gbps and to EDGE at 1~Gbps, with $6\times$ acceleration for ML workloads.
\end{itemize}
Inter-DU links are set to 1~Gbps (limited lateral connectivity) and self-loops carry effectively infinite bandwidth to model zero-cost co-located communication. Table~III summarizes the link structure. This topology captures the core dispersed-computing tension: the Cloud is $6\times$ faster in compute but reachable only via low-bandwidth backhaul, while DUs are slowest but enjoy 100~Gbps fronthaul to EDGE.

\begin{table}[t]
  \centering
  \caption{Network topology parameters.}
  \label{tab:topology-params}
  \begin{tabular}{l c l l}
    \hline
    Link Type & BW & Unit & Example \\
    \hline
    Fronthaul & 100 & Gbps & DU $\leftrightarrow$ EDGE \\
    EDGE--RIC & 50 & Gbps & EDGE $\leftrightarrow$ RIC \\
    RIC--Cloud & 10 & Gbps & RIC $\leftrightarrow$ Cloud \\
    DU--RIC (E2) & 1 & Gbps & DU $\leftrightarrow$ RIC \\
    DU--Cloud & 1 & Gbps & DU $\leftrightarrow$ Cloud \\
    Inter-DU & 1 & Gbps & DU $\leftrightarrow$ DU \\
    EDGE--Cloud & 1 & Gbps & EDGE $\leftrightarrow$ Cloud \\
    Self-loop & $\infty$ & Gbps & (co-located) \\
    \hline
  \end{tabular}
\end{table}

\subsection{Stage 3: Scheduling with Extended SAGA}
We use the SAGA framework~[11] as our scheduling engine, which provides modular implementations of HEFT~[6], MCT, MinMin, MaxMin, and Duplex. Given the profiled DAG and network topology, SAGA constructs the expected execution time matrix $W$ (where $w_{i,v}=c(t_i)/s(v)$) and the communication time matrix $C$ (where $c_{i,j,u,v}=d(e_{i,j})/\mathrm{bw}(u,v)$), then applies the chosen heuristic to produce a task-to-node mapping with start and end times.

We extend SAGA\footnote{Extended SAGA: \url{https://github.com/ANRGUSC/saga/tree/feature/pinning}\label{fn:saga}} with a node-pinning constraint that restricts designated tasks to specific compute nodes. In the slicing scheduler, T0 (Collect State) is pinned to its originating DU because it aggregates local RAN telemetry---buffer status reports and platform KPIs---that exists only at that physical site. T2 (Link Quality Map) is similarly pinned because it processes sounding reference signals from the local physical layer. Pinning is implemented by setting $w_{i,v}=\infty$ for all non-target nodes, ensuring the scheduler's greedy selection never assigns the task elsewhere. This extension does not exist in upstream SAGA and is a necessary adaptation for O-RAN workloads where physical-layer telemetry is inherently localized.

When the DAG is replicated across $N_{\mathrm{cells}}$ cells, the scheduler processes all $N_{\mathrm{cells}}\times 8$ tasks jointly. Each cell's T0 and T2 are pinned to their respective DU, while the remaining tasks compete for placement on the shared EDGE, RIC, and Cloud nodes. This joint scheduling naturally captures cross-cell contention for shared infrastructure---a critical effect absent from single-cell analyses. The scheduler outputs a JSON file that serves as input to the simulation stage.

\subsection{Stage 4: MintEDGE DAG Simulation}
The final stage validates the schedule under 5G network conditions. We extend MintEDGE~[12] with a custom DAG simulation module that replays the scheduler's task-to-node mapping while enforcing dependency constraints, link bandwidth sharing, and transport delays.

Given the schedule JSON, the simulator initializes each task with its assigned node and scheduled start time. At each step, it checks whether all predecessors of a ready task have completed and whether input data has been delivered. The actual start time may differ from the SAGA-predicted time due to bandwidth contention: when multiple data transfers share a link simultaneously, effective bandwidth per transfer is reduced proportionally. This contention modeling, absent from HEFT's idealized cost model, is precisely what motivates the simulation stage.

For each task, the simulator decomposes execution into three phases, inspired by Dapper's span-based tracing~[22]: (1) uplink time---transferring input data from predecessor nodes to the task's node, with parallel transfers sharing ingress bandwidth; (2) compute time---the instruction count divided by the node's speed factor and a configurable base compute speed; and (3) downlink time---delivering outputs to successor nodes.

Upon completion, the simulator verifies two correctness properties: (a) all DAG dependencies are satisfied and (b) the timing decomposition is faithful---uplink + compute + downlink = actual duration for every task. The simulated makespan typically exceeds the HEFT-predicted makespan because contention is now modeled; the magnitude of this gap is itself informative, indicating whether contention-aware heuristics~[7] should be considered. The simulator outputs per-task metrics (start/end times, phase breakdown) and aggregate metrics (makespan, per-node utilization), enabling the latency analysis in Section~VI.

\section{O-DAG Framework}
We evaluate our framework through a systematic parameter study that varies three independent dimensions of the O-RAN deployment: the number of user equipment (UEs), the number of cells, and the number of network slices. For each configuration, we execute the full four-stage pipeline (Section~IV) and compare five scheduling algorithms at the SAGA stage, followed by MintEDGE simulation of the best schedule. This section describes the experimental setup, the parameter space, the profiled DAG costs across configurations, and the metrics used for evaluation.

\subsection{Baseline Configuration}
The baseline configuration uses 7 cells (DU0--DU6), 5 network slices, and 10{,}000 UEs. This represents a moderate-scale suburban deployment where each DU serves approximately 1{,}400 UEs distributed across 5 slices. The slicing scheduler DAG is replicated once per cell, yielding $7\times 8=56$ tasks scheduled jointly across 10 compute nodes (7 DUs + EDGE + RIC + Cloud). The network topology follows the specification in Section~IV-B with speed factors $s(\mathrm{DU})=1$, $s(\mathrm{EDGE})=s(\mathrm{RIC})=3$, $s(\mathrm{Cloud})=6$.

\subsection{Parameter Space}
We vary one parameter at a time while holding the other two at their baseline values (10{,}000 UE, 7 Cells, 5 Slices).

UE count (5{,}000 / 10{,}000 / 50{,}000) scales the per-task instruction count proportionally, modeling increased computational load as more users generate buffer status reports and sounding reference signals. At 5{,}000 UEs, the DAG is compute-light and communication costs may dominate; at 50{,}000 UEs, ML inference tasks (T1, T2) become significantly heavier, potentially shifting the critical path and favoring cloud placement for its $6\times$ speed advantage.

Cell count (2 / 7 / 20) scales the number of independent DAG instances and the number of DU nodes in the topology. At 2 cells, cross-cell contention for EDGE/RIC/Cloud is minimal (16 tasks, 2 DUs). At 20 cells, 160 tasks compete for 3 shared aggregation nodes, stressing the scheduler’s ability to balance load and exposing contention on shared fronthaul and backhaul links.

Slice count (2 / 5 / 10) scales the width of the inter-slice scheduling problem within each cell. Increasing slices from 2 to 10 increases the data volume processed by T3 (Inter Slice Budget) and T5 (Intra Slice Schedule), as more slices require more PRB allocation decisions. This dimension tests whether the scheduling decision remains stable as the per-task workload grows within a fixed topology.

\subsection{Profiled DAG Costs Across Configurations}
Before scheduling, each configuration is independently profiled by DagProfiler (Stage~1) to capture the actual per-task instruction counts and per-edge communication volumes. Table~IV reports the aggregate compute and communication costs across all seven configurations, and Table~V provides the full per-task instruction breakdown.

\begin{table}[t]
  \centering
  \caption{Aggregate profiled DAG costs per configuration.}
  \label{tab:profiled-agg-costs}
  \begin{tabular}{l r r c}
    \hline
    Configuration & Total Instr. & Total Bits & CCR \\
    \hline
    Baseline (10K/7/5) & 12.72M & 12.49M & 1.02 \\
    UE = 5K & 6.45M & 6.25M & 1.03 \\
    UE = 20K & 62.84M & 62.41M & 1.01 \\
    Cells = 2 & 12.53M & 12.48M & 1.00 \\
    Cells = 20 & 13.19M & 12.52M & 1.05 \\
    Slices = 2 & 12.61M & 12.49M & 1.01 \\
    Slices = 10 & 12.89M & 12.51M & 1.03 \\
    \hline
  \end{tabular}
\end{table}

The compute-to-communication ratio (CCR = total instructions / total bits) remains remarkably stable near 1.0 across all configurations, indicating that the slicing scheduler DAG is balanced between compute and communication costs. This balance is significant: it means that neither purely compute-aware nor purely communication-aware scheduling heuristics are universally favored, and the optimal placement depends on the fine-grained interaction between task costs and link bandwidths.

\begin{table}[t]
  \centering
  \caption{Task instruction counts (in thousands) across configurations.}
  \label{tab:profiled-per-task}
  \begingroup
  \setlength{\tabcolsep}{3pt}
  \footnotesize
  \begin{tabular}{l r r r r r r r}
    \hline
    Task & Base & 5K(UE) & 20K(UE) & 2C & 20C & 2S & 10S \\
    \hline
    T0 & 3{,}760 & 1{,}880 & 18{,}800 & 3{,}710 & 3{,}890 & 3{,}760 & 3{,}760 \\
    T1 & 54 & 39 & 174 & 37 & 99 & 40 & 78 \\
    T2 & 7{,}720 & 3{,}860 & 38{,}600 & 7{,}720 & 7{,}720 & 7{,}720 & 7{,}720 \\
    T3 & 55 & 55 & 55 & 16 & 159 & 22 & 111 \\
    T4 & 480 & 240 & 2{,}400 & 480 & 480 & 480 & 480 \\
    T5 & 116 & 111 & 156 & 40 & 313 & 58 & 212 \\
    T6 & 510 & 205 & 2{,}050 & 410 & 410 & 410 & 410 \\
    T7 & 120 & 60 & 600 & 121 & 120 & 120 & 120 \\
    \hline
  \end{tabular}
  \endgroup
\end{table}

Three distinct scaling patterns emerge from Table~V. First, UE count scales almost all tasks proportionally: T0 (state collection), T2 (link quality map), T4 (UE weight builder), and T6 (apply and simulate) all exhibit near-linear scaling with UE count (e.g., T0: 1.88M at 5K $\rightarrow$ 3.76M at 10K $\rightarrow$ 18.8M at 20K). This is expected, as each of these tasks processes per-UE data structures whose sizes grow linearly. Critically, T3 (Inter Slice Budget) remains invariant to UE count at 55K instructions---its computation depends only on the number of slices, not the number of users per slice.

Second, cell count has minimal effect on the per-cell compute cost (total: 12.53M at 2 cells vs. 13.19M at 20 cells, a 5.3\% increase) but selectively scales the coordination tasks T1, T3, and T5. T3 (Inter Slice Budget) grows 10$\times$ from 16K to 159K instructions as the number of neighboring cells increases from 2 to 20, reflecting the growing coordination overhead for cross-cell interference management. Similarly, T5 (Intra Slice Schedule) grows from 40K to 313K. The critical-path tasks T0 and T2 remain effectively unchanged, explaining why the overall makespan in Section~VI grows only modestly with cell count.

Third, slice count selectively scales T3 and T5 while leaving all other tasks unchanged. T3 grows from 22K at 2 slices to 111K at 10 slices (5$\times$), and T5 from 58K to 212K (3.7$\times$), reflecting the additional budgeting and scheduling decisions required per slice. Notably, T0 and T2 are completely invariant to slice count---state collection and link quality estimation operate at the physical layer, independent of the number of logical slices. The edge communication costs (Table~IV, total bits column) are similarly invariant across cell and slice variations, confirming that the data payload sizes are determined primarily by the UE count.

These profiling patterns directly predict the scheduling behavior observed in Section~VI: UE count determines the overall makespan magnitude, while cell and slice counts govern the degree of cross-cell contention and the workload balance across tasks.

\subsection{Scheduling Algorithms}
At Stage~3, we evaluate five scheduling algorithms from SAGA~[11], all extended with our node-pinning constraint (Section~IV-C):
\begin{itemize}
  \item \textbf{HEFT}~[6] ranks tasks by upward rank and greedily assigns each to the processor minimizing earliest finish time (EFT), using an insertion-based policy that exploits idle slots.
  \item \textbf{MCT}~[23] (Minimum Completion Time) assigns each task to the processor that yields the minimum completion time, without upward-rank prioritization.
  \item \textbf{MinMin}~[24] considers all unscheduled tasks, computes each task’s minimum completion time across all processors, and schedules the task with the globally smallest minimum first.
  \item \textbf{MaxMin}~[24] follows the same structure as MinMin but schedules the task with the globally largest minimum completion time first, favoring critical-path tasks.
  \item \textbf{Duplex}~[24] combines MinMin and MaxMin strategies and selects the schedule with the lower makespan.
\end{itemize}
For each experimental configuration, all five schedulers are run on the same profiled DAG and topology. The best-performing scheduler’s schedule is then forwarded to the MintEDGE simulation stage for validation.

\subsection{Evaluation Metrics}
We report the following metrics for each experiment:
\begin{itemize}
  \item \textbf{SAGA makespan:} the predicted end-to-end completion time (in instruction-time units) produced by each scheduling algorithm, assuming sequential per-node execution and no link contention. This is the standard metric for comparing scheduling heuristics~[6], [11].
  \item \textbf{Simulated makespan:} the actual completion time produced by the MintEDGE DAG simulation, which accounts for concurrent per-node execution, bandwidth contention, and explicit uplink transfer phases.
  \item \textbf{SAGA--simulation gap:} the percentage difference between the SAGA-predicted and simulated makespans. This metric quantifies the aggregate effect of the execution-model differences between SAGA and MintEDGE (detailed in Section~VI-A) and serves as a compact diagnostic for modeling mismatch.
  \item \textbf{Per-task latency decomposition:} for the simulated schedule, the breakdown of each task’s duration into uplink, compute, and downlink components. Aggregating these across all tasks reveals whether the deployment is compute-bound, communication-bound, or contention-bound across the pipeline.
\end{itemize}

\section{Results and Discussion}
We present results along three experimental axes---UE count, cell count, and slice count---evaluating both the SAGA scheduling stage and the MintEDGE simulation stage. All experiments use the profiled slicing scheduler DAG and network topology described in Sections~III--IV. Before examining individual results, we first characterize the key modeling difference between SAGA and MintEDGE that governs the interpretation of all SAGA--simulation gaps.

\subsection{Execution Model Differences}
The SAGA--simulation gap arises from two fundamental differences in how SAGA and MintEDGE model task execution.
\begin{itemize}
  \item \textbf{Per-node concurrency.} SAGA enforces strictly sequential execution within each compute node. It searches for non-overlapping time slots between tasks already assigned to that node, meaning no two tasks on the same node may execute concurrently. MintEDGE, by contrast, allows parallel execution on the same node---tasks are dispatched as soon as all DAG predecessors have completed and input data has arrived, regardless of whether other tasks are already running on that node.
  \item \textbf{Transfer timing.} SAGA models communication as an abstract cost derived from data volume divided by link bandwidth, embedded into each task’s earliest-start-time calculation as a single scalar. MintEDGE explicitly simulates the uplink transfer phase: input data from each predecessor is transmitted over the network link with bandwidth sharing when multiple transfers compete for the same link. This more granular transfer model can either increase or decrease the effective communication delay relative to SAGA’s estimate, depending on link contention.
\end{itemize}
These two effects act in opposing directions. The concurrency effect consistently decreases the simulated makespan relative to HEFT (tasks overlap rather than serialize). The transfer-timing effect can increase it when uplink contention is significant. The sign and magnitude of the observed gap reflect the balance between these two forces across different workload regimes, as we demonstrate below.

\subsection{Effect of UE Count}
\begin{table}[t]
  \centering
  \caption{SAGA makespan by scheduler under varying UE count (7 cells, 5 slices).}
  \label{tab:ue-saga}
  \begin{tabular}{l r r r l}
    \hline
    Sched. & 5K UEs & 10K UEs & 50K UEs & $\Delta$ \\
    \hline
    HEFT & 6{,}212{,}237 & 12{,}353{,}832 & 61{,}418{,}673 & Best \\
    MinMin & +0.04\% & +0.10\% & +0.28\% & 2--3 \\
    Duplex & +0.04\% & +0.10\% & +0.28\% & 2--3 \\
    MCT & +0.48\% & +0.14\% & +0.45\% & 4--5 \\
    MaxMin & +0.99\% & +0.53\% & +0.44\% & 4--5 \\
    \hline
  \end{tabular}
\end{table}

Table~VI reports the SAGA-predicted makespans. HEFT produces the shortest makespan at all three UE counts, with MinMin and Duplex within 0.04--0.28\% and MCT/MaxMin trailing by up to 0.99\%. The makespan scales nearly linearly with UE count: the 5K$\rightarrow$10K ratio is 1.99$\times$ and the 10K$\rightarrow$50K ratio is 4.97$\times$, confirming that computation (which scales with UE count) dominates the critical path over communication (which does not). The tight clustering of all five schedulers (spread $<1$\% at every UE level) indicates that the DAG’s fork--join structure with pinned entry tasks leaves limited room for algorithmic differentiation.

\begin{table}[t]
  \centering
  \caption{MintEDGE simulation vs.\ HEFT prediction under varying UE count.}
  \label{tab:ue-sim}
  \begin{tabular}{l r r r}
    \hline
    UEs & HEFT Predicted & Simulated & Gap \\
    \hline
    5K & 6{,}212{,}237 & 6{,}105{,}340 & $-1.72$\% \\
    10K & 12{,}353{,}832 & 12{,}175{,}340 & $-1.44$\% \\
    50K & 61{,}418{,}673 & 60{,}735{,}340 & $-1.11$\% \\
    \hline
  \end{tabular}
\end{table}

Table~VII compares the HEFT-predicted makespan with the MintEDGE-simulated makespan. The simulated makespan is consistently lower, with gaps of $-1.72$\%, $-1.44$\%, and $-1.11$\% at 5K, 10K, and 50K UEs respectively. This negative gap is explained by the per-node concurrency difference: HEFT serializes all tasks assigned to the same node, whereas MintEDGE executes them in parallel as soon as predecessor constraints are satisfied. With 7 cells and 56 tasks competing for 10 nodes, HEFT stacks multiple tasks sequentially on popular nodes (particularly Cloud and EDGE), while MintEDGE overlaps their execution.

The gap narrows as UE count increases (from $-1.72$\% at 5K to $-1.11$\% at 50K). At higher UE counts, each individual task takes longer to execute, so the compute phase dominates the per-task duration. When compute time dwarfs communication and scheduling overhead, the benefit of parallel dispatch diminishes---the critical-path task on the bottleneck node determines the makespan regardless of whether other tasks overlap. In the limiting case of a single infinitely long task, HEFT’s serial model and MintEDGE’s parallel model converge.

\subsection{Effect of Cell Count}
\begin{table}[t]
  \centering
  \caption{SAGA makespan by scheduler under varying cell count (10K UEs, 5 slices).}
  \label{tab:cells-saga}
  \begin{tabular}{l r r r l}
    \hline
    Sched. & 2 Cells & 7 Cells & 20 Cells & $\Delta$ \\
    \hline
    HEFT & 12{,}253{,}398 & 12{,}353{,}832 & 12{,}932{,}775 & Best \\
    MinMin & +0.17\% & +0.10\% & +0.77\% & 2$\rightarrow$5 \\
    Duplex & +0.17\% & +0.10\% & +0.02\% & 2$\rightarrow$2 \\
    MaxMin & +0.56\% & +0.53\% & +0.02\% & 4$\rightarrow$2 \\
    MCT & +0.98\% & +0.14\% & +1.29\% & 5 \\
    \hline
  \end{tabular}
\end{table}

Table~VIII reveals that the HEFT makespan increases only modestly from 2 to 20 cells (12.25M $\rightarrow$ 12.93M, a 5.5\% increase), despite a 10$\times$ increase in task count (16 $\rightarrow$ 160 tasks). This sublinear growth demonstrates that HEFT distributes load effectively: per-cell pinned tasks (T0, T2) absorb most of the additional workload on their respective DU nodes, while non-pinned tasks are spread across the shared infrastructure.

A notable shift occurs in the scheduler ranking at 20 cells: MaxMin and Duplex rise to within 0.02\% of HEFT, while MinMin degrades to +0.77\% and MCT to +1.29\%. This reversal---MinMin outperformed MaxMin at 2 and 7 cells but not at 20---aligns with SAGA’s PISA analysis~[11]: at higher contention, MaxMin’s strategy of prioritizing the longest task first better handles the competition for shared EDGE/RIC/Cloud resources, whereas MinMin’s shortest-first policy greedily fills fast nodes with small tasks, leaving heavy tasks queued.

\begin{table}[t]
  \centering
  \caption{MintEDGE simulation vs.\ HEFT prediction under varying cell count.}
  \label{tab:cells-sim}
  \begin{tabular}{l r r r}
    \hline
    Cells & HEFT Predicted & Simulated & Gap \\
    \hline
    2 & 12{,}253{,}398 & 12{,}100{,}115 & $-1.25$\% \\
    7 & 12{,}353{,}832 & 12{,}175{,}340 & $-1.44$\% \\
    20 & 12{,}932{,}775 & 12{,}932{,}775 & 0.00\% \\
    \hline
  \end{tabular}
\end{table}

The simulation results (Table~IX) show a convergence. At 2 and 7 cells, the concurrency effect dominates: MintEDGE runs tasks in parallel on shared nodes where HEFT serializes them, producing gaps of $-1.25$\% and $-1.44$\%. At 20 cells, however, the gap collapses to exactly 0.00\%. With 160 tasks distributed across 10 nodes, the schedule becomes so densely packed that HEFT’s sequential model is effectively correct---every node is fully utilized with back-to-back tasks, leaving no room for parallel overlap. Simultaneously, the increased number of inter-node transfers at 20 cells raises uplink contention in MintEDGE’s transfer model, counteracting any residual concurrency benefit. The two modeling differences (per-node parallelism favoring MintEDGE, uplink contention penalizing it) cancel precisely at this operating point.

\subsection{Effect of Slice Count}
\begin{table}[t]
  \centering
  \caption{SAGA makespan by scheduler under varying slice count (7 cells, 10K UEs).}
  \label{tab:slices-saga}
  \begin{tabular}{l r r r l}
    \hline
    Sched. & 2 Slices & 5 Slices & 10 Slices & $\Delta$ \\
    \hline
    HEFT & 12{,}292{,}728 & 12{,}353{,}832 & 12{,}418{,}047 & Best \\
    MinMin & +0.32\% & +0.10\% & +4.87\% & 2$\rightarrow$5 \\
    Duplex & +0.32\% & +0.10\% & +0.91\% & 2$\rightarrow$3 \\
    MaxMin & +0.85\% & +0.53\% & +0.91\% & 4$\rightarrow$2 \\
    MCT & +0.51\% & +0.14\% & +4.58\% & 3$\rightarrow$4 \\
    \hline
  \end{tabular}
\end{table}

Table~X shows the most algorithmic differentiation in our study. At 2 and 5 slices, all schedulers remain within 0.85\% of HEFT. At 10 slices, however, MinMin and MCT degrade sharply to +4.87\% and +4.58\% above HEFT, respectively, while MaxMin and Duplex remain within 0.91\%. This divergence reveals a fundamental sensitivity: as the per-task workload grows (more slices increase the data processed by T3 and T5), MinMin’s greedy shortest-first policy assigns too many lightweight tasks to fast nodes early, leaving heavy tasks stranded on slower nodes. MaxMin’s longest-first strategy avoids this trap by prioritizing critical-path tasks. This finding is consistent with SAGA’s adversarial analysis~[11], which showed that schedulers appearing equivalent on benchmarks can diverge dramatically on specific workload structures.

\begin{table}[t]
  \centering
  \caption{MintEDGE simulation vs.\ HEFT prediction under varying slice count.}
  \label{tab:slices-sim}
  \begin{tabular}{l r r r}
    \hline
    Slices & HEFT Predicted & Simulated & Gap \\
    \hline
    2 & 12{,}292{,}728 & 12{,}370{,}925 & +0.64\% \\
    5 & 12{,}353{,}832 & 12{,}175{,}340 & $-1.44$\% \\
    10 & 12{,}418{,}047 & 12{,}370{,}925 & $-0.38$\% \\
    \hline
  \end{tabular}
\end{table}

The simulation results (Table~XI) exhibit a qualitative shift: at 2 slices, the simulated makespan exceeds the HEFT prediction by +0.64\%---the only positive gap in our entire study. This sign reversal is explained by the balance of the two execution-model differences. With only 2 slices, per-task compute costs are low, so the concurrency benefit (parallel execution on shared nodes) is small. At the same time, the transfer-timing effect becomes relatively more significant: MintEDGE explicitly models uplink transfer phases with bandwidth sharing, and at low compute loads these transfer delays constitute a larger fraction of end-to-end latency. The uplink overhead outweighs the concurrency benefit, producing a net positive gap.

At 5 slices, compute costs are higher and the concurrency benefit reasserts dominance ($-1.44$\%). At 10 slices, the gap narrows to $-0.38$\%: the per-task workload is now large enough that the bottleneck node’s critical-path task determines makespan regardless of parallelism, similar to the convergence observed with increasing UE count. The sign change in the gap is a key diagnostic: it identifies a transition from a transfer-dominated regime (few slices, where MintEDGE’s uplink model penalizes performance beyond HEFT’s estimate) to a compute-dominated regime (many slices, where HEFT’s serial per-node model overestimates).

\subsection{Summary of Findings}
Three principal findings emerge from the experimental evaluation:
\begin{enumerate}
  \item \textbf{HEFT is consistently optimal for this DAG.} Across all configurations, HEFT produces the lowest SAGA makespan. The upward-rank prioritization combined with insertion-based scheduling is well-suited to the slicing scheduler’s fork--join topology with pinned entry tasks. However, the margin of superiority varies: $<1$\% in most configurations, but up to 4.87\% over MinMin at 10 slices.
  \item \textbf{Scheduler rankings are workload-dependent.} The relative ordering of the five algorithms changes across configurations. Most notably, MinMin and MCT degrade at high slice counts and high cell counts, while MaxMin improves. This confirms SAGA’s adversarial finding~[11] that benchmark-level comparisons are insufficient---algorithm selection must be informed by the target workload’s structure.
  \item \textbf{The execution-model gap diagnoses workload regimes.} The SAGA--simulation gap reflects the interplay between two modeling differences: SAGA’s sequential per-node execution versus MintEDGE’s explicit uplink transfer modeling. Negative gaps indicate compute-dominated regimes where the concurrency effect dominates; the sole positive gap (+0.64\% at 2 slices) identifies a transfer-dominated regime where uplink overhead exceeds the benefit of parallel dispatch. Convergence to 0.00\% at 20 cells indicates that high node utilization eliminates both effects.
\end{enumerate}

\section{Conclusion}
\sloppy
We presented O-DAG, an end-to-end framework that bridges four previously disconnected domains—O-RAN architecture, DAG profiling, heterogeneous scheduling, and cellular simulation—into a single reproducible pipeline. Starting from a concrete RAN slicing scheduler, we formalized an eight-task DAG, profiled it using our open-source DagProfiler tool, scheduled it across a three-tier dispersed topology using an extended SAGA framework with node-pinning constraints, and validated the resulting placement via a custom DAG simulation module built on MintEDGE\footnote{Extended MintEDGE: https://github.com/ANRGUSC/MintEDGE}.

Several directions remain for future work. First, extending DagProfiler to capture workload-dependent cost distributions would improve scheduling fidelity beyond the current assumption of homogeneous per-UE instruction costs. Second, the node-pinning constraint could be generalized to soft constraints with migration penalties, enabling the scheduler to trade off data locality against load balance. Third, cross-validation on a hardware testbed would establish the correspondence between our simulation-based results and protocol-level behavior. Finally, applying the framework to additional O-RAN applications would test the generalizability of both the profiling schema and the scheduling recommendations. All O-DAG artifacts are publicly available to support reproducibility.

\bibliographystyle{IEEEtran}
% Avoid stale/invalid generated .bbl causing “missing \item” in thebibliography.
% Force BibTeX to (re)generate the bibliography from refs.bib.
\nocite{*}
\bibliography{refs}

% Generated by IEEEtran.bst, version: 1.14 (2015/08/26)
\begin{thebibliography}{10}
\providecommand{\url}[1]{#1}
\csname url@samestyle\endcsname
\providecommand{\newblock}{\relax}
\providecommand{\bibinfo}[2]{#2}
\providecommand{\BIBentrySTDinterwordspacing}{\spaceskip=0pt\relax}
\providecommand{\BIBentryALTinterwordstretchfactor}{4}
\providecommand{\BIBentryALTinterwordspacing}{\spaceskip=\fontdimen2\font plus
\BIBentryALTinterwordstretchfactor\fontdimen3\font minus \fontdimen4\font\relax}
\providecommand{\BIBforeignlanguage}[2]{{%
\expandafter\ifx\csname l@#1\endcsname\relax
\typeout{** WARNING: IEEEtran.bst: No hyphenation pattern has been}%
\typeout{** loaded for the language `#1'. Using the pattern for}%
\typeout{** the default language instead.}%
\else
\language=\csname l@#1\endcsname
\fi
#2}}
\providecommand{\BIBdecl}{\relax}
\BIBdecl

\bibitem{ORAN2018WhitePaper}
{O-RAN ALLIANCE}, ``{O-RAN: Towards an Open and Smart RAN},'' O-RAN ALLIANCE, White Paper, 10 2018.

\bibitem{ETSI2024TS103982}
{ETSI}, ``{TS} 103 982 (v8.0.0),'' ETSI, Tech. Rep., 01 2024.

\bibitem{NGO2024680}
M.~V. Ngo \emph{et~al.}, ``{RAN} intelligent controller ({RIC}): From open-source implementation to real-world validation,'' \emph{ICT Express}, vol.~10, no.~3, pp. 680--691, 2024.

\bibitem{Doro2022DApps}
S.~D'Oro \emph{et~al.}, ``{dApps}: Distributed applications for real-time inference and control in {O-RAN},'' \emph{IEEE Communications Magazine}, vol.~60, no.~11, 2022.

\bibitem{Ananthanarayanan2025OpenRAN}
G.~Ananthanarayanan \emph{et~al.}, ``{Distributed AI platform for the 6G RAN},'' in \emph{Proceedings of the ACM OpenRAN '25}, 2025.

\bibitem{Topcuoglu2002HEFT}
H.~Topcuoglu \emph{et~al.}, ``Performance-effective and low-complexity task scheduling for heterogeneous computing,'' \emph{IEEE Transactions on Parallel and Distributed Systems}, vol.~13, no.~3, 2002.

\bibitem{Wu2022Contention}
Q.~Wu \emph{et~al.}, ``Endpoint communication contention-aware cloud workflow scheduling,'' \emph{IEEE Transactions on Automation Science and Engineering}, vol.~19, no.~2, 2022.

\bibitem{Chen2016Uncertainty}
H.~Chen \emph{et~al.}, ``Uncertainty-aware real-time workflow scheduling in the cloud,'' in \emph{Proceedings of IEEE IPDPS}, 2016.

\bibitem{Genez2017Robust}
T.~A.~L. Genez \emph{et~al.}, ``A robust scheduler for workflow ensembles under bandwidth uncertainties,'' in \emph{Proceedings of IEEE CLOUD}, 2017.

\bibitem{Viramontes2024DIME}
R.~Viramontes and A.~Davoodi, ``{DIME: Distributed inference model estimation},'' in \emph{Proceedings of IEEE SMARTCOMP}, 2024.

\bibitem{Coleman2024Adversarial}
J.~Coleman and B.~Krishnamachari, ``{PISA: An Adversarial Approach to Comparing Task Graph Scheduling Algorithms},'' in \emph{2025 IEEE International Parallel and Distributed Processing Symposium (IPDPS)}.\hskip 1em plus 0.5em minus 0.4em\relax IEEE, 2025, pp. 54--66.

\bibitem{Gomez2023MintEDGE}
B.~G{\'o}mez \emph{et~al.}, ``{MintEDGE: Multi-tier sImulator for eNergy-aware sTrategies in Edge Computing},'' in \emph{Proceedings of ACM MobiCom '23}, 2023.

\bibitem{Polese2024Colosseum}
M.~Polese \emph{et~al.}, ``{Colosseum: The open RAN digital twin},'' \emph{IEEE Open Journal of the Communications Society}, vol.~5, 2024.

\bibitem{Schmidt2021FlexRIC}
R.~Schmidt \emph{et~al.}, ``{FlexRIC: An SDK for next-generation SD-RANs},'' in \emph{Proceedings of ACM CoNEXT '21}, 2021.

\bibitem{6676705}
D.~Kliazovich \emph{et~al.}, ``{CA-DAG: Communication-Aware Directed Acyclic Graphs for Modeling Cloud Computing Applications},'' in \emph{2013 IEEE Sixth International Conference on Cloud Computing}, 2013, pp. 277--284.

\bibitem{Coleman2024Parametric}
J.~Coleman \emph{et~al.}, ``{Evaluating the Impact of Algorithmic Components on Task Graph Scheduling},'' in \emph{Workshop on Job Scheduling Strategies for Parallel Processing}.\hskip 1em plus 0.5em minus 0.4em\relax Springer, 2025, pp. 243--262.

\bibitem{Yi2017LAVEA}
S.~Yi \emph{et~al.}, ``Lavea: Latency-aware video analytics on edge computing platform,'' in \emph{Proceedings of the Second ACM/IEEE Symposium on Edge Computing (SEC)}, 10 2017.

\bibitem{8737478}
W.~Zhang \emph{et~al.}, ``{Hetero-edge}: Orchestration of real-time vision applications on heterogeneous edge clouds,'' in \emph{IEEE INFOCOM 2019 - IEEE Conference on Computer Communications}, 2019, pp. 1270--1278.

\bibitem{Naouri2021MEC}
A.~Naouri \emph{et~al.}, ``A novel framework for mobile-edge computing by optimizing task offloading,'' \emph{IEEE Internet of Things Journal}, vol.~8, no.~16, pp. 13\,065--13\,076, 08 2021.

\bibitem{Li2024MTEC}
X.~Li \emph{et~al.}, ``Dynamic {DAG}-application scheduling for multi-tier edge computing in heterogeneous networks,'' 09 2024, arXiv preprint arXiv:2409.10839.

\bibitem{Queiroz2024Offload}
G.~F.~C. de~Queiroz \emph{et~al.}, ``A flexible algorithm to offload {DAG} applications for edge computing,'' \emph{Journal of Network and Computer Applications}, vol. 222, 2024.

\bibitem{Sigelman2010Dapper}
B.~H. Sigelman \emph{et~al.}, ``Dapper, a large-scale distributed systems tracing infrastructure,'' Google, Tech. Report, 2010.

\bibitem{Armstrong1998RelativePerformance}
R.~Armstrong \emph{et~al.}, ``The relative performance of various mapping algorithms is independent of sizable variances in run-time predictions,'' in \emph{Proceedings Seventh Heterogeneous Computing Workshop (HCW'98)}, 1998, pp. 79--87.

\bibitem{Braun2001ElevenStaticHeuristics}
T.~D. Braun \emph{et~al.}, ``A comparison of eleven static heuristics for mapping a class of independent tasks onto heterogeneous distributed computing systems,'' \emph{Journal of Parallel and Distributed Computing}, vol.~61, no.~6, pp. 810--837, 2001.

\end{thebibliography}

\end{document}